# High Current Density 2D/3D Esaki Tunnel Diodes

*Sriram Krishnamoorthy[†]\*, Edwin W. Lee II[†], Choong Hee Lee[†], Yuewei Zhang[†], William D. McCulloch[§], Jared M. Johnson[‡], Jinwoo Hwang[‡], Yiying Wu[§], and Siddharth Rajan[†‡]\**

[†]Department of Electrical and Computer Engineering, [‡]Department of Materials Science and Engineering, and [§]Department of Chemistry and Biochemistry, The Ohio State University, Columbus, OH 43210, United States

ABSTRACT

**The integration of two-dimensional materials such as transition metal dichalcogenides with bulk semiconductors offer interesting opportunities for 2D/3D heterojunction-based novel device structures without any constraints of lattice matching. By exploiting the favorable band alignment at the GaN/MoS$_2$ heterojunction, an Esaki interband tunnel diode is demonstrated by transferring large area, Nb-doped, p-type MoS$_2$ onto heavily n-doped GaN. A peak current density of 446 A/cm$^2$ with repeatable room temperature negative differential resistance, peak to valley current ratio of 1.2, and minimal hysteresis was measured in the MoS$_2$/GaN non-epitaxial tunnel diode. A high current density of 1 kA/cm$^2$ was measured in the Zener mode (reverse bias) at -1 V bias. The GaN/MoS$_2$ tunnel junction was also modeled by treating MoS$_2$ as a bulk semiconductor, and the electrostatics at the**

**2D/3D interface was found to be crucial in explaining the experimentally observed device characteristics.**

Two-dimensional (2D) layered materials such as transition metal dichalcogenides(TMDs)[1] with no out-of-plane bonds offer interesting opportunities for integration with bulk semiconductors such as Si, GaAs and GaN. Integration of diverse bulk semiconductors is limited by the need for lattice matching and extended defect control. With no constraints of lattice matching, 2D materials can be integrated with bulk semiconductors to create novel vertical and lateral heterostructures. More recently, researchers have started exploring such 2D/3D heterojunctions for device applications. $MoS_2$/Si heterojunctions have been investigated for device applications such as photodetectors[2,3,4] and solar cells[5]. Researchers have also studied $MoS_2$/GaAs[6,7], $MoS_2$/InP[8], $MoS_2$/SiC[9] and $MoS_2$/GaN[10,11,12,13] heterojunctions for a range of device applications . 2D/2D ($MoS_2$/$WSe_2$[14,15,16], BP/$SnSe_2$[17], $WSe_2$/$SnSe_2$[18] ) and 2D/3D ($MoS_2$/Ge)[19] heterojunctions-based interband tunnel diodes and tunneling field effect transistors have also been experimentally demonstrated. 2D/3D heterojunctions are in particular very attractive for low power logic devices with steep subthreshold slope[19] as 2D channel material enables excellent electrostatic gate control, and the use of 3D semiconductor enables ultra-sharp doping profiles and engineering of the band lineup to increase tunneling current.

Integration of 2D-layered materials with wide band gap semiconductors such as GaN is particularly interesting due to the complimentary properties of these material systems. While GaN and AlN have high breakdown fields enabling high power/voltage device operation, there are fundamental challenges in p-type doping of GaN, AlN[20] and ZnO[21]. Transition metal dichalcogenides with lower band gap can offer excellent current spreading properties due to high mobility even when they are atomically thin. Previous investigations on GaN/$MoS_2$ PN diodes[10] indicated a very low value of conduction band offset ($\Delta E_c \sim 0.2$ eV). The very small value of conduction band offset resulted in a space charge region in GaN with a built-in potential ($V_{bi}$ = 1.5 V) much smaller than the band gap of GaN (3.4 eV)[10]. Such a band line up is expected to be suitable for an Esaki interband tunnel diode[22] provided the depletion width is designed to be sufficiently narrow by employing heavy doping. In this work, we report a $p^+$ $MoS_2$/ $n^+$ GaN interband tunnel diode formed by transferring multilayer Nb-doped $p^+$ $MoS_2$ to molecular beam epitaxy grown $n^+$ GaN. We observe high forward and reverse tunneling current with the characteristic negative differential resistance observed in forward bias at room temperature. Minimal hysteresis measured in the IV characteristics indicates the promise of such non-epitaxial heterointerfaces for extreme band engineering without lattice matching constraints.

$p^+$ $MoS_2$ was grown by sulfurizing a sputtered Mo/Nb/Mo stack on sapphire ( 1 cm x 0.8 cm) with $MoS_2$ powder as the sulfur source in a vacuum sealed quartz tube at $1100^o$ C. Nb acts as p-type dopant in $MoS_2$ [23,24] and we have reported large area heavily p-type doped multilayer $MoS_2$ on sapphire[25] using this approach[26]. Large-area multilayer, p-type $MoS_2$ was characterized by X-ray diffraction(XRD) and the films were found to be crystalline as evident from (002), (004) and (006) $MoS_2$ peaks observed in XRD spectra (Figure 1(a)). Atomic force microscopy images (Fig. 1(b)) revealed a smooth surface morphology with a root mean square roughness of 0.6 nm. P-type conductivity was confirmed in the as-grown $p^+$ $MoS_2$/sapphire film using Hall measurements with Ni/Au contact pads e-beam evaporated at the four corners of the film. A sheet resistance of 164 k$\Omega$/□, sheet charge density of 7.6 x $10^{13}$ $cm^{-2}$ and a Hall mobility of 0.5 $cm^2$/ V s was measured. The equivalent volume charge density is calculated to be 1.1 x $10^{20}$ $cm^{-3}$ from the measured sheet charge density and thickness of the film (7 nm). Ni/Au/Ni metal contacts were also patterned on large area $p^+$ $MoS_2$ using optical lithography. Using a transfer length measurement structure (see Supplementary Figure S1), the resistance was measured as a function of length between contact pads (5 um to 25 um). From the linear fit of the total resistance versus length, a sheet resistance value of 154 kΩ/□ was extracted. The sheet resistance value extracted from transfer length measurement structure is in close agreement with the value extracted from large area hall measurements, indicating uniform conductivity across the large area p-MoS$_2$ sample.

$n^+$ GaN (20 nm) was grown on n-GaN/Sapphire templates using N$_2$ plasma-assisted molecular beam epitaxy (MBE) in Ga-rich growth conditions[27] at a substrate temperature of 700$^o$C using standard effusion cells for gallium and silicon dopant. A high silicon doping density of 5 x 10$^{19}$ cm$^{-3}$ was used to enhance interband tunneling. The target doping density was achieved using SIMS measurements on calibration samples (not shown here). The Ga-rich growth conditions resulted in excess Ga metal droplets on the surface, which were removed using hydrochloric acid.

$p^+$ MoS$_2$ was transferred onto MBE-grown $n^+$ GaN using a benign DI water-based process[28] reported earlier[29]. A PDMS stamp was pressed on to $p^+$ MoS$_2$/ sapphire and detached to create an air gap at the interface. The sample was then immersed in DI water. The MoS$_2$ layer was lifted off from the substrate and mm-sized fragments of MoS$_2$ films floated in DI water. $p^+$ MoS$_2$ was then fished out using MBE-grown $n^+$ GaN/sapphire sample. The sample was then dried overnight and annealed in a vacuum chamber at 110 $^o$C to remove any water molecules trapped at the MoS$_2$/GaN interface. The resulting device stack is shown in Fig. 2 (a). An optical microscope image of the layer-transferred MoS$_2$/GaN diode is shown in Fig. 2 (b), showing the Ni/Au/Ni metal contacts on continuous MoS$_2$ layers. Raman spectra (Renishaw, 514 nm laser) of as-grown $p^+$ MoS$_2$/sapphire, $n^+$ GaN/sapphire and layer transferred $p^+$ MoS$_2$/ $n^+$ GaN/sapphire is shown in Fig. 2 (c). The characteristic $E^1_{2g}$ and $A_{1g}$ vibrational modes were observed with a peak separation of 26 cm$^{-1}$ were observed and a peak intensity ratio of 1.16 in as-grown p$^+$MoS$_2$/sapphire. Epitaxial GaN/sapphire showed a very strong Raman peak at 570 cm$^{-1}$ corresponding to the $E_2$ mode in GaN and a faint $A_{1g}$ mode (419 cm$^{-1}$) corresponding to c-plane sapphire. Both the characteristic $E^1_{2g}$ and $A_{1g}$ modes of MoS$_2$ (peak ratio of 1.08) as well as the $E_2$ mode of GaN were observed in the layer-transferred MoS$_2$/GaN sample. However, the signal to noise ratio in the MoS$_2$/GaN was about one third of the as-grown MoS$_2$/sapphire indicating degradation of material quality during layer transfer. Inductively coupled plasma/reactive ion etching (ICP-RIE) with BCl$_3$/Ar–based chemistry was used to etch MoS$_2$ using Ni/Au/Ni metal stack as the etch mask for device isolation. Large area Ni/Au/Ni metal pads directly in contact with n$^+$ GaN (regions with no MoS$_2$) was used as ohmic contacts to n$^+$ GaN.

A schematic of the equilibrium band diagram of the p$^+$ MoS$_2$/ n$^+$ GaN tunnel diode is shown in Fig. 2(d). Here, the experimentally measured conduction band offset (~ 0.2 eV) between MoS$_2$ and GaN from a previous report is used[10]. It should be noted that there is a van der Waals gap at the MoS$_2$/ GaN interface and a van der Waals gap between each S-Mo-S layers of MoS$_2$. The qualitative charge profile in the MoS$_2$/GaN TJ is also shown in Figure 2(d). The space charge regions due to depletion of donors in GaN, depletion of acceptors in MoS$_2$, negative spontaneous polarization[30] sheet charge at the surface of GaN, charged surface donors[31] and unintentional positive sheet charge due to O, Si contaminants[32] at the GaN surface play an important role in the electrostatics of the MoS$_2$/GaN junction. In addition to the sources of charge mentioned above, the MoS$_2$/GaN band offset is expected to be different from the electron affinity difference due to the presence of interfacial dipoles[33]. Heavy doping on both sides of the junction can considerably reduce the depletion region width. The total positive sheet charge due to surface states and O,Si contaminants is denoted as $N_{interface, net}^+$. Under forward bias (Fig. 2 (e)), electrons in the conduction band ($\Gamma$ valley) of GaN tunnel into empty states (holes) available in the valence band of multilayer MoS$_2$ ($\Gamma$ valley). Although multilayer MoS$_2$ is an indirect gap semiconductor, the interband tunneling process between the $\Gamma$ valleys is expected to be akin to a direct tunneling process. With increasing forward bias, the tunneling current is expected to increase until the conduction and valence band are out of alignment. Beyond this forward bias, the tunneling current drop until forward bias diffusion current becomes the dominant conduction mechanism. This results in negative differential resistance under forward bias. Under reverse bias (Fig. 2(f)), the electrons in the valence band of MoS$_2$ tunnel into empty states in the conduction band of GaN, and this is referred to as Zener tunneling.

Z-contrast scanning transmission electron microscope (STEM) images of GaN/MoS$_2$ tunnel diode device is shown in Fig.3. Individual 2D layers of MoS$_2$, the van der Waals gaps, and the individual Mo and S atoms could be resolved in the STEM images. In comparison to previous reports of oxidized interfaces between 2D material and bulk semiconductors[19], the heterojunction between layer-transferred MoS$_2$ and GaN reported in this work is fairly abrupt in nature. The basal plane spacing ('a' lattice spacing) measured from the STEM images indicated alignment of the basal planes of GaN and MoS$_2$.

Electrical characteristics of p$^+$ MoS$_2$/ n$^+$ GaN tunnel diodes are shown in Fig. 4. Under reverse bias, a very high current density was measured even in low reverse bias regime. A reverse current density of 1 kA/cm$^2$ (315 A/cm$^2$) was measured at a reverse bias of 1 V (0.5 V). Under forward bias, negative differential resistance was observed at room temperature, which is a characteristic feature of interband tunneling in Esaki diodes. A peak current density of 446 A/cm$^2$ was measured at a forward bias voltage of 0.8 V. A valley current of 368 A/cm$^2$ was measured at a voltage of 1.2 V, with a peak to valley current ratio of 1.2. Negative differential resistance was found to be repeatable and robust during multiple cycles of measurements. The double hump observed in the negative differential resistance regime (inset to Fig. 4 (b)) is attributed to the oscillations in the measurement circuit due to the presence of negative resistance and is a measurement artifact[34].

The device exhibited minimal hysteresis in its IV characteristics when the voltage sweep direction was reversed (Fig.4 (c)). The minimal hysteresis observed in a layer-transferred, non-epitaxial heterojunction such as this indicates the possibility of making high quality 2D/3D heterojunctions with minimal charge trapping for a wide range of device applications.

SILVACO ATLAS[35] was used to model $p^+$ $MoS_2$/GaN TJ. Multilayer $MoS_2$ was treated as a bulk 3D semiconductor for simulation purposes, and a conduction band offset of 0.2 eV was used based on previous experimental results from $MoS_2$/GaN heterojunctions[10]. A hole effective mass of $0.5996m_0$[36] was used. Holes are populated in the Γ valley of the valence band in $p^+$ $MoS_2$, and hence the tunneling process is expected to be direct. As such, the direct band gap value of $MoS_2$ (1.2 eV) was used for the simulation. The following parameters were used in the simulation: Conduction band effective density of states (DOS)[35] of 6.222e+018 cm$^{-3}$, Valence band effective DOS of 6.516e+018 cm$^{-3}$ and Electron affinity[37] of 3.8 eV. Band structure was computed in SILVACO using k.p method and tunneling current was computed using non-local band to band tunneling model.

Tunneling current density was computed at room temperature for different values of $N_{interface, net}^+$. The peak current density calculated in the case of $N_{interface, net}^+ = 4 \times 10^{13}$ cm$^{-2}$ matches the experimentally observed peak current density (denoted as star in Figures 5 (c)). It should be noted that the calculated current density is "intrinsic" tunneling current. In reality, series and contact resistances increases the total voltage drop in the device. The total voltage drop in the device can be computed as $V_{EXPT} = V_{SIMULATON,TJ} + J_{SIMULATION,TJ} * \rho_{series}$. Using $\rho_{series} = 1.7 \times 10^{-3}$ $\Omega cm^2$, and the computed I-V curve for $N_{interface, net}^+ = 4 \times 10^{13}$ $cm^{-2}$, the model described here was successfully able to match the experimental peak voltage, forward and reverse bias current density (Fig. 5 (d)). This indicates that the electrostatics at the 2D/3D heterojunction, interface sheet charge and polarization charge in this case, play a crucial role in determining the electrical characteristics of the 2D/3D heterojunction.

The low resistance GaN/MoS$_2$ tunnel diodes reported here can enable a number of novel device applications. With controlled p-type doping, n-type doping and gating, steep slope tunneling FETs can be realized with GaN/MoS$_2$ tunnel diodes. The 2D/3D tunnel junction reported here, in conjunction with 2D/2D tunnel junctions can enable series connection of multiple atomically thin 2D heterostructure-based active regions for applications such as multi junction solar cells, multiple color light emitting diodes and multi-spectral photodetectors. With no constraints of lattice matching, vertical heterostructures with extreme band gaps can be integrated and connected in series with such tunnel junctions. Traditionally interband tunnel junctions have been realized in epitaxial heterostructures. This work demonstrates the potential of non-epitaxial heterostructures for high performance device applications.

In summary, we report high current density 2D/3D tunnel junctions with non-epitaxial layer-transferred p+ MoS$_2$ on to n$^+$ GaN. Repeatable, room temperature negative differential resistance with a peak current of 446 A/cm$^2$, peak to valley ratio of 1.2 and minimal hysteresis is reported. Such 2D/3D heterojunction-based tunnel diodes can enable novel high performance vertical heterostructures based on 2D-layered materials and bulk semiconductors.

FIGURES

Table of Contents Figure

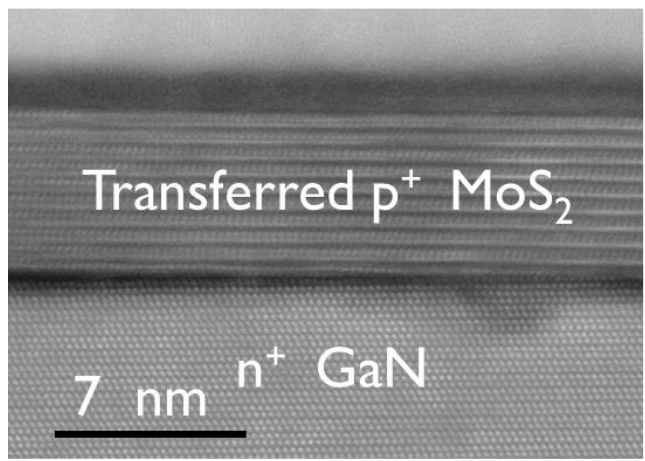 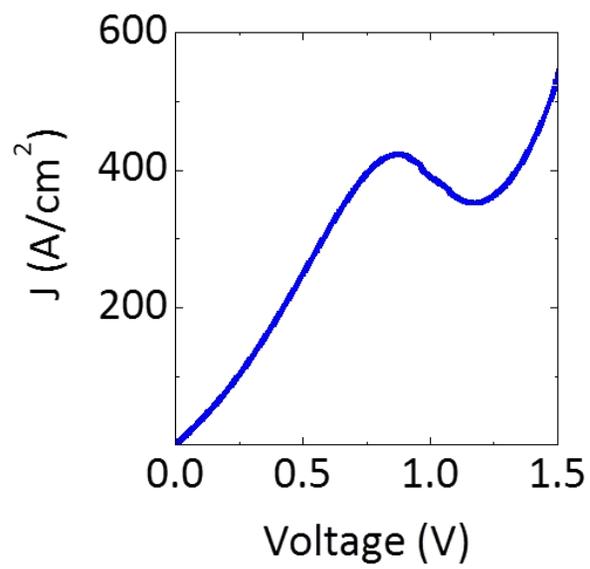

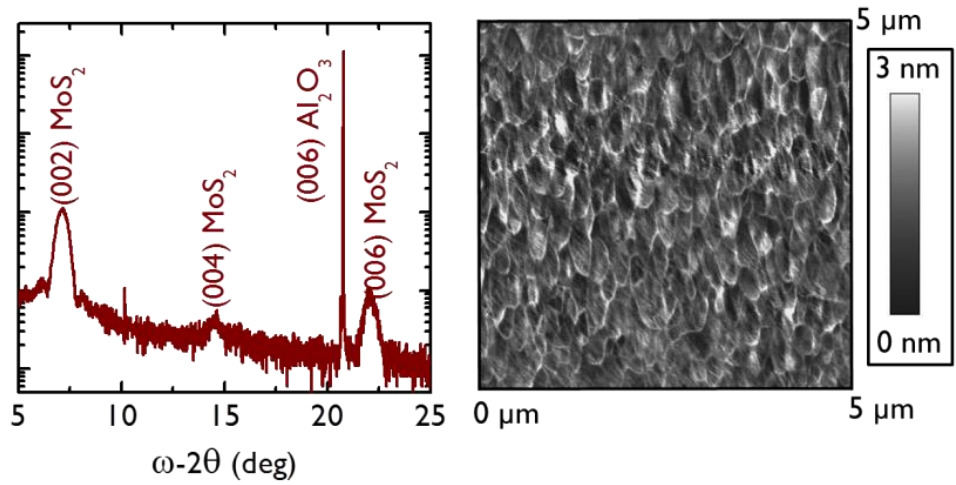

**Figure 1.** (a) X-ray diffraction data of crystalline large area p$^+$ MoS$_2$ grown by chalcogenization of sputtered Mo/Nb/Mo stack on sapphire substrates (b) Atomic force microscope image of p$^+$ MoS$_2$ indicating smooth surface morphology (RMS roughness = 0.6 nm).

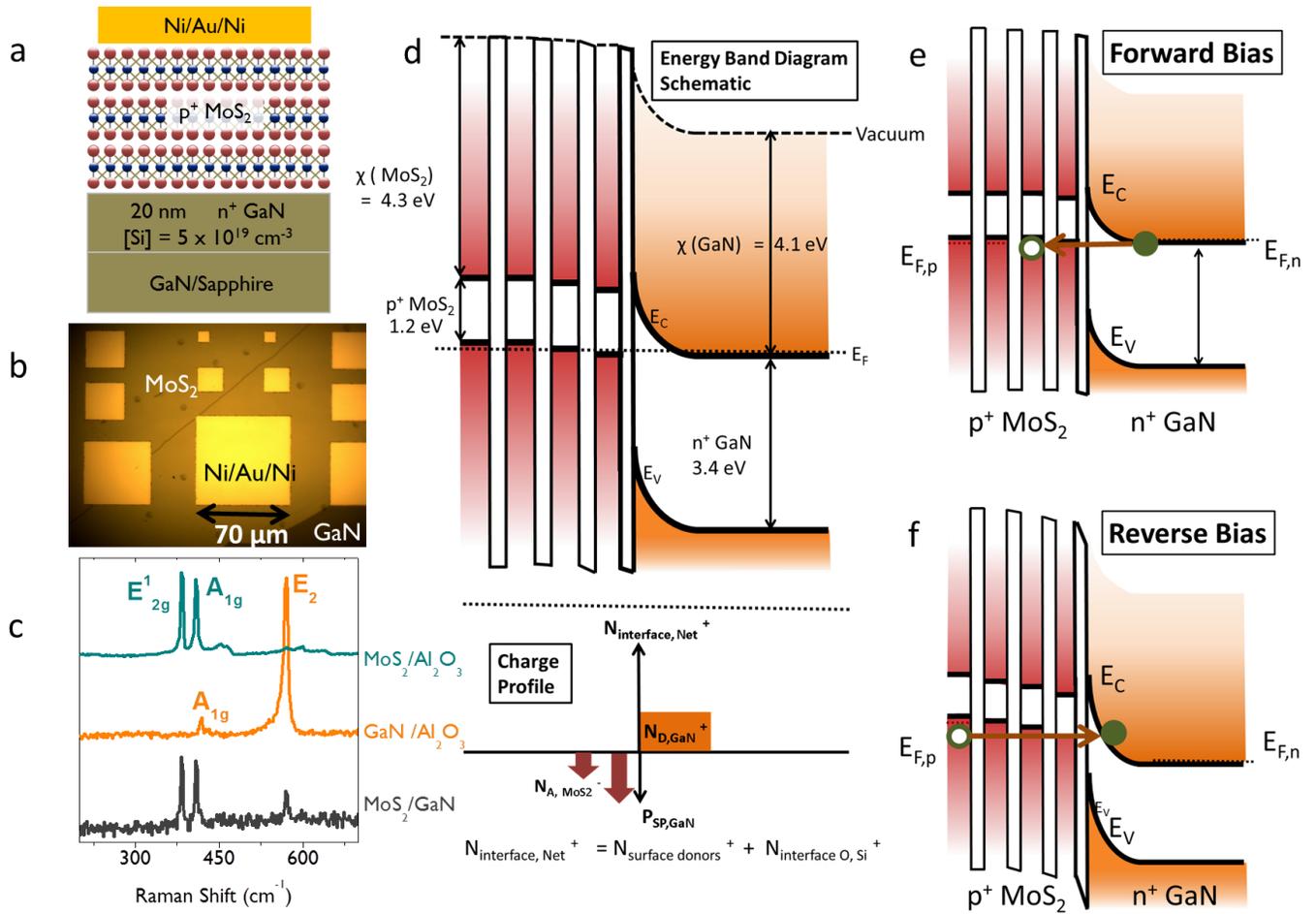

**Figure 2.** (a) Schematic of $p^+$ $MoS_2$/ $n^+$ GaN device structure, (b) Optical micrograph image of $MoS_2$/GaN tunnel diode with Ni/Au/Ni ohmic contact to $p^+$ $MoS_2$. Ni/Au/Ni is used as a etch mask to etch $MoS_2$ to realize device isolation. (c) Raman spectra of as-grown $MoS_2$/sapphire, MBE-grown $n^+$ GaN/sapphire and layer-transferred $MoS_2$/GaN tunnel diode showing characteristic Raman peaks. (d) Equilibrium energy band diagram and charge diagram schematic showing various components of charge contribution to the junction electrostatics. (e) Schematic band diagram under forward bias and (f) reverse bias illustrating the interband tunneling process.

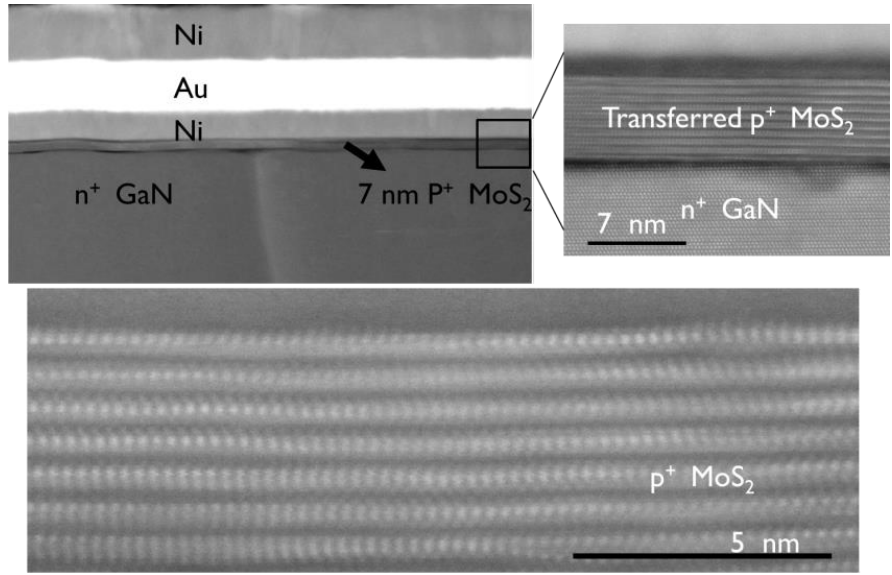

**Figure 3.** Scanning transmission electron microscope images of layer transferred $p^+$ $MoS_2$/ $n^+$ GaN tunnel diode device showing continuous $MoS_2$ coverage, abrupt interface, individual Mo, S atoms and van der Waals gaps.

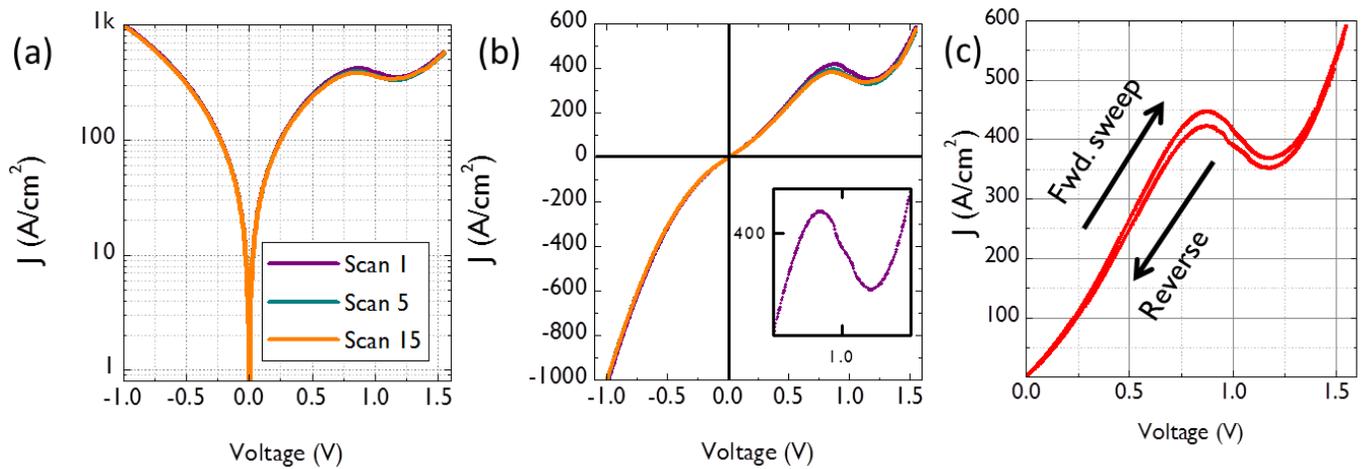

**Figure 4.** (a) Linear, (b) log current-voltage characteristics of $p^+$ $MoS_2$/ $n^+$ GaN tunnel diode showing repeatable room temperature negative differential resistance, and (c) minimal hysteresis measured in between forward and reverse sweeps.

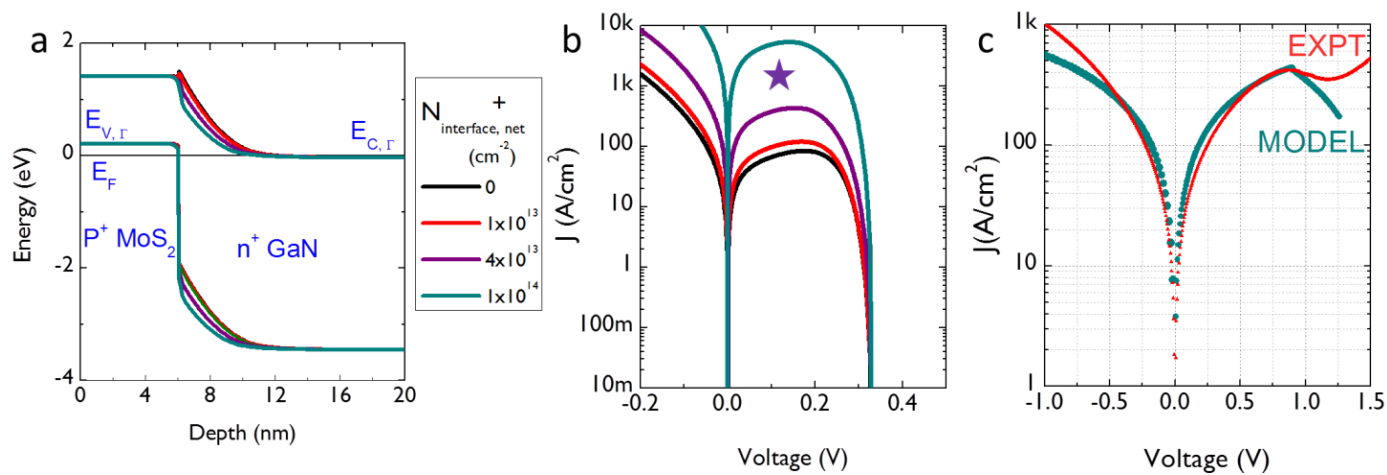

**Figure 5.** (a) Equilibrium band diagram, (b) Log J-V characteristics of the "intrinsic" tunnel diode as a function of net interface positive charge ($N_{interface,net}^+$). The calculated peak current density value matches with the experimentally measured value for $N_{interface,\,net}^+ = 4E13$ cm$^{-2}$. (d) With the addition of series resistance, the simulated IV characterisitcs match with the experimentally measured characteristics.

ASSOCIATED CONTENT

**Supporting Information**. Discussion of sheet resistance measurement in p$^+$ MoS$_2$, Surface morphology of MBE-grown GaN/Sapphire, layer transferred MoS$_2$ on GaN, Effect of n-type doping in MoS$_2$/GaN TJ. This material is available free of charge via the Internet at http://pubs.acs.org.

AUTHOR INFORMATION

**Corresponding Author**

* Email: krishnamoorthy.13@osu.edu; rajan@ece.osu.eduACKNOWLEDGMENT

The authors would like to acknowledge the support from Air Force Office of Scientific Research (Program manager: Dr. Kenneth Goretta) under Contract No. FA9550-15-1-0294.REFERENCES

(1) Radisavljevic, B.; Radenovic, A.; Brivio, J.; Giacometti, V.; Kis, A. Single-layer $MoS_2$ transistors. *Nature nanotechnology* **2011**, *6*(3), 147-150.

(2) Esmaeili-Rad, Mohammad R., and Sayeef Salahuddin. "High performance molybdenum disulfide amorphous silicon heterojunction photodetector." *Scientific reports* 3 (2013).

(3) Wang, Liu, Jiansheng Jie, Zhibin Shao, Qing Zhang, Xiaohong Zhang, Yuming Wang, Zheng Sun, and Shuit-Tong Lee. "MoS2/Si Heterojunction with Vertically Standing Layered Structure for Ultrafast, High-Detectivity, Self-Driven Visible–Near Infrared Photodetectors." *Advanced Functional Materials* 25, no. 19 (2015): 2910-2919.

(4) Li, Bo, Gang Shi, Sidong Lei, Yongmin He, Weilu Gao, Yongji Gong, Gonglan Ye et al. "3D Band Diagram and Photoexcitation of 2D–3D Semiconductor Heterojunctions." *Nano letters* 15, no. 9 (2015): 5919-5925.

(5) Hao, Lanzhong, Yunjie Liu, Wei Gao, Zhide Han, Qingzhong Xue, Huizhong Zeng, Zhipeng Wu, Jun Zhu, and Wanli Zhang. "Electrical and photovoltaic characteristics of MoS2/Si pn junctions." *Journal of Applied Physics* 117, no. 11 (2015): 114502.

(6) Lin, Shisheng, Xiaoqiang Li, Peng Wang, Zhijuan Xu, Shengjiao Zhang, Huikai Zhong, Zhiqian Wu, Wenli Xu, and Hongsheng Chen. "Interface designed MoS2/GaAs heterostructure solar cell with sandwich stacked hexagonal boron nitride." *Scientific reports* 5 (2015).

(7) Xu, Zhijuan, Shisheng Lin, Xiaoqiang Li, Shengjiao Zhang, Zhiqian Wu, Wenli Xu, Yanghua Lu, and Sen Xu. "Monolayer MoS2/GaAs heterostructure self-driven photodetector with extremely high detectivity." *arXiv preprint arXiv:1512.06867* (2015).

(8) Lin, Shisheng, Peng Wang, Xiaoqiang Li, Zhiqian Wu, Zhijuan Xu, Shengjiao Zhang, and Wenli Xu. "Gate tunable monolayer MoS2/InP heterostructure solar cells." *Applied Physics Letters* 107, no. 15 (2015): 153904.

(9) Lee II, Edwin W., Lu Ma, Digbijoy N. Nath, Choong Hee Lee, Aaron Arehart, Yiying Wu, and Siddharth Rajan. "Growth and electrical characterization of two-dimensional layered MoS2/SiC heterojunctions." *Applied Physics Letters* 105, no. 20 (2014): 203504.

(10) Lee II, Edwin W., Choong Hee Lee, Pran K. Paul, Lu Ma, William D. McCulloch, Sriram Krishnamoorthy, Yiying Wu, Aaron R. Arehart, and Siddharth Rajan. "Layer-transferred MoS2/GaN PN diodes." *Applied Physics Letters* **2015**, 107, 10,103505.

(11) Ruzmetov, Dmitry, Kehao Zhang, Gheorghe Stan, Berc Kalanyan, Ganesh R. Bhimanapati, Sarah M. Eichfeld, Robert A. Burke et al. "Vertical 2D/3D Semiconductor Heterostructures Based on Epitaxial Molybdenum Disulfide and Gallium Nitride." *ACS nano* **2016**.

(12) Liao, Jiamin, Baisheng Sa, Jian Zhou, Rajeev Ahuja, and Zhimei Sun. "Design of high-efficiency visible-light photocatalysts for water splitting: MoS2/AlN (GaN) heterostructures." *The Journal of Physical Chemistry C* 118, no. 31 (**2014**): 17594-17599.

(13) Jeong, Hyun, Seungho Bang, Hye Min Oh, Hyeon Jun Jeong, Sung-Jin An, Gang Hee Han, Hyun Kim et al. "Semiconductor–Insulator–Semiconductor Diode Consisting of Monolayer MoS2, h-BN, and GaN Heterostructure." *ACS nano* 9, no. 10 (**2015**): 10032-10038.

(14) Lee, C.H.; Lee, G.H.; Van Der Zande, A.M.; Chen, W.; Li, Y.; Han, M.; Cui, X.; Arefe, G.; Nuckolls, C.; Heinz, T.F.; Guo, J.; Hone, J.; Kim, P. Atomically thin p–n junctions with van der Waals heterointerfaces. *Nature nanotechnology* **2014**, *9*(9), pp.676-681.

(15) Roy, T.; Tosun, M.; Cao, X.; Fang, H.; Lien, D. H.; Zhao, P.; Chen, Y.Z.; Chueh, Y.L.; Guo, J; Javey, A. Dual-Gated MoS2/WSe2 van der waals tunnel diodes and transistors. *ACS Nano* **2015**, *9*(2), 2071-2079.

(16) Nourbakhsh, A.; Zubair, A.; Dresselhaus, M.S.; Palacios, T. Transport Properties of a $MoS_2$/$WSe_2$ Heterojunction Transistor and its Potential for Application. *Nano letters* **2016**, 16 (2), pp 1359–1366

(17) Yan, R.; Fathipour, S.; Han, Y.; Song, B.; Xiao, S.; Li, M.; Ma, N.; Protasenko, V.; Muller, D.A.; Jena, D.; Xing, H.G. Esaki diodes in van der Waals heterojunctions with broken-gap energy band alignment. *Nano letters* **2015**, *15*(9), pp.5791-5798.

(18) Roy, T.; Tosun, M.; Hettick, M.; Ahn, G.H.; Hu, C.; Javey, A. 2D-2D tunneling field-effect transistors using $WSe_2$/$SnSe_2$ heterostructures. *Applied Physics Letters* **2016**, *108*(8), p.083111.

(19) Sarkar, D.; Xie, X.; Liu, W.; Cao, W.; Kang, J.; Gong, Y.; Kraemer, S.; Ajayan, P.M.; Banerjee, K. A subthermionic tunnel field-effect transistor with an atomically thin channel. *Nature* **2015**, *526*(7571), pp.91-95.

(20) Tanaka, T., A. Watanabe, H. Amano, Y. Kobayashi, I. Akasaki, S. Yamazaki, and M. Koike. "p-type conduction in Mg-doped GaN and Al0. 08Ga0. 92N grown by metalorganic vapor phase epitaxy." *Applied physics letters* 65, no. 5 (**1994**): 593-594.

(21) Park, C. H., S. B. Zhang, and Su-Huai Wei. "Origin of p-type doping difficulty in ZnO: The impurity perspective." *Physical Review B* 66, no. 7 (2002): 073202.

(22) Esaki, L. New phenomenon in narrow germanium p− n junctions. *Physical review* **1958**, *109*(2), p.603.

(23) J. A. Wilson and A. D. Yoffe, Adv. Phys. 18(73), 193–335 (1969)

(24) R. S. Title and M. W. Shafer, Phys. Rev. Lett. **28**(13), 808–810 (1972)

(25) Ma, L.;, Nath , D.N.; Lee II ,E.W.; Lee, C.H.; Yu,M.; Arehart, A.; Rajan, S.; Wu, Y. Epitaxial growth of large area single-crystalline few-layer MoS2 with high space charge mobility of 192 cm$^2$ V$^{-1}$ s$^{-1}$." *Applied Physics Letters* 105, no. 7 (**2014**): 072105.

(26) Laskar, Masihhur R., Digbijoy N. Nath, Lu Ma, Edwin W. Lee II, Choong Hee Lee, Thomas Kent, Zihao Yang et al. "p-type doping of MoS2 thin films using Nb." *Applied Physics Letters* 104, no. 9 (**2014**) : 092104.

(27) Tarsa, E. J., B. Heying, X. H. Wu, P. Fini, S. P. DenBaars, and J. S. Speck. "Homoepitaxial growth of GaN under Ga-stable and N-stable conditions by plasma-assisted molecular beam epitaxy." *Journal of applied physics* 82, no. 11 (**1997**): 5472-5479.

(28) Gurarslan, A.; Yu, Y.; Su, L.; Yu, Y.; Suarez, F.; Yao, S.; Zhu, Y.; Ozturk, M.; Zhang, Y.; Cao, L. Surface-energy-assisted perfect transfer of centimeter-scale monolayer and few-layer MoS2 films onto arbitrary substrates. *ACS nano* **2014**, *8*(11), pp.11522-11528.

(29) Lee, Choong Hee, William McCulloch, Edwin W. Lee II, Lu Ma, Sriram Krishnamoorthy, Jinwoo Hwang, Yiying Wu, and Siddharth Rajan. "Transferred large area single crystal MoS2 field effect transistors." *Applied Physics Letters* 107, no. 19 (2015): 193503.

(30) Bernardini, F.; Fiorentini, V. ; Vanderbilt, D. Spontaneous polarization and piezoelectric constants of III-V nitrides. *Physical Review B* **1997**, *56*(16), p.R10024.

(31) Ibbetson, J.P.; Fini, P.T.; Ness, K.D.; DenBaars, S.P.; Speck, J.S.; Mishra, U.K. Polarization effects, surface states, and the source of electrons in AlGaN/GaN heterostructure field effect transistors. *Applied Physics Letters* **2000**, *77*(2), pp.250-252.

(32) Lee, W.; Ryou, J.H.; Yoo, D.; Limb, J.; Dupuis, R.D.; Hanser, D.; Preble, E.; Williams, N.M. ; Evans, K. Optimization of Fe doping at the regrowth interface of GaN for applications to III-nitride-based heterostructure field-effect transistors. *Applied Physics Letters* **2007**, *90*(9), p.093509.

(33) Schlaf, R.; Lang, O.; Pettenkofer, C.; Jaegermann, W. Band lineup of layered semiconductor heterointerfaces prepared by van der Waals epitaxy: Charge transfer correction term for the electron affinity rule. *Journal of applied physics* **1999**, *85*(5), pp.2732-2753.

(34) Bao, M.; Wang, K.L. Accurately measuring current-voltage characteristics of tunnel diodes. *IEEE Transactions on Electron Devices,* **2006**, *53*(10), pp.2564-2568.

(35) http://www.silvaco.com/products/tcad/device_simulation/atlas/atlas.html

(36) Liu, L., Kumar, S.B., Ouyang, Y. and Guo, J., 2011. Performance limits of monolayer transition metal dichalcogenide transistors. Electron Devices, IEEE Transactions on, 58(9), pp.3042-3047.

(37) Lee, K., Kim, H.Y., Lotya, M., Coleman, J.N., Kim, G.T. and Duesberg, G.S., 2011. Electrical characteristics of molybdenum disulfide flakes produced by liquid exfoliation. Advanced Materials, 23(36), pp.4178-4182.